# Effects of Nerve Bundle Geometry on Neurotrauma Evaluation

Ilaria Cinelli, Michel Destrade, Peter McHugh and Maeve Duffy

*Abstract*— *Objective:* We confirm that alteration of a neuron structure can induce abnormalities in signal propagation for nervous systems, as observed in brain damage. Here, we investigate the effects of geometrical changes and damage of a neuron structure in two scaled nerve bundle models, made of myelinated nerve fibres or unmyelinated nerve fibres. *Methods:* We propose a 3D finite element model of nerve bundles, combining a real-time full electro-mechanical coupling, a modulated threshold for spiking activation and independent alteration of the electrical properties for each fibre. We then simulate mechanical compression and tension to induce damage at the membrane of a nerve bundle made of four fibres. We examine the resulting changes in strain and neural activity by considering in turn the cases of intact and traumatized nerve membranes. *Results:* Our results show lower strain and lower electrophysiological impairments in unmyelinated fibres than in myelinated fibres, higher deformation levels in larger bundles, and higher electrophysiological impairments in smaller bundles. *Conclusion:* We conclude that the insulation sheath of myelin constricts the membrane deformation and scatters plastic strains within the bundle; that larger bundles deform more than small bundles; and that small fibres tolerate a higher level of elongation before mechanical failure.

## I. Introduction

Head injuries may result in Traumatic Brain Injury (TBI), which is categorized as mild, moderate and severe based on clinical symptoms and post-mortem histology (Hemphill, Dauth, Yu, Dabiri, & Parker, 2015; Hosmane et al., 2011; Ma, Zhang, Wang, & Chen, 2016; Wright & Ramesh, 2012; Y. P. Zhang et al., 2014). The rapid acceleration-deceleration of the head during TBI generates a diffusive form of microscale damage, such as Diffuse Axonal Injury (DAI) and microvascular damage (Hemphill et al., 2015; Ma et al., 2016; Wright & Ramesh, 2012; Y. P. Zhang et al., 2014). However, microscale damage is difficult to detect with the current medical imaging technology due to haemorrhages, hematomas and tissue lesions of the neighbouring injured area (Hemphill et al., 2015; Kan, Ling, & Lu, 2012; Wright & Ramesh, 2012). This difficulty increases the risk of developing future neurodegenerative disease (Hemphill et al., 2015; Kan et al., 2012).

Previous biomechanical studies of brain injuries have highlighted the importance of brain microenvironment and of neural tissue responses in the understanding of disease pathogenesis initiated by TBI (Hemphill et al., 2015). They established that tensile axonal strain is the most realistic mechanism for generating DAI at the cellular level (Allison C Bain & Meaney, 2000;

Cinelli, Destrade, Mchugh, & Duffy, 2017; Jérusalem, García-Grajales, Merchán-Pérez, & Peña, 2014; Wright & Ramesh, 2012).

At the next level, the distribution of diffuse damage is found to be non-uniform throughout the brain tissue, suggesting that tissue regions and cellular structures within the brain are affected differently (Hemphill et al., 2015). Tissue heterogeneity has a significant influence on the mechano-transduction of mechanical forces into physiological and neural responses of nervous cells (Hemphill et al., 2015), and therefore models that accurately account for tissue structure are needed for an effective modelling of damage.

Experiments have also revealed a close link between changes in electrical signal propagation and changes in the geometrical structure of neurons (P.-C. Zhang, Keleshian, & Sachs, 2001). Indeed, a geometrical alteration of neural morphology can modify the propagation properties of the action potential, for instance by delaying propagation (Boucher, Joós, & Morris, 2012; Cinelli, Destrade, Duffy, & McHugh, 2017c; Mohagheghian, 2015). A detailed investigation of non-recoverable deformations of the neural microenvironment (injuries (Jérusalem et al., 2014; Wright & Ramesh, 2012), trauma (Jérusalem et al., 2014), tumours (Mohagheghian, 2015)) is needed to evaluate and estimate the role of nerve bundle geometry in changing neural activity.

Recent progress in physiological measurements has led to new insights into damaged neuronal behaviour, where electrophysiological and functional deficits of the neural activity are known to be functions of the applied strain and strain rate (Boucher et al., 2012; Geddes, Cargill, & LaPlaca, 2003; Jérusalem et al., 2014). Electrophysiological impairments (such as leaking ionic channels (Boucher et al., 2012; Yu, Morris, Joós, & Longtin, 2012)) are associated with structural damage of the neuron tissue. The loss of nerve membrane integrity due to an applied deformation leads to changes in electrical signal propagation (Galbraith, Thibault, & Matteson, 1993; Yu et al., 2012). Furthermore, injury pathologies in nerve fibres are also initiated and influenced by strain and strain rate, which have a significant impact on the time of neural death and pathomorphology, respectively (Bar-kochba, Scimone, Estrada, & Franck, 2016). For instance, experimental studies on human axons show that morphological changes may tolerate dynamic stretch at strains up to 65% (Smith, Wolf, Lusardi, Lee, & Meaney, 1999), manifesting both an elastic recovery and a delayed elastic response along the fibre length (Smith et al., 1999).

Here we evaluate the influence of neuron morphology in neurotrauma, which refers to the alteration of neural activity in a mechanical-injured nerve (Galbraith et al., 1993; Geddes et al., 2003), by using a fully coupled electro-mechanical model in the finite element (FE) software package Abaqus. Our purpose is to evaluate strain distributions leading to neurotrauma in damaged nerve bundles of different types and sizes during signalling. This work aims at improving the understanding the mechano-transduction of mechanical loads below the threshold for mechanical failure, on neural responses in nerve bundles and fibres.

In contrast with previous modelling efforts (Jérusalem et al., 2014; Mohagheghian, 2015), we propose a fully coupled 3D electro-mechanical model of a nerve bundle (Cinelli, Destrade, Duffy, & McHugh, 2017b; Cinelli, Destrade, Mchugh, et al., 2017), which includes electro-mechanical coupling (Alvarez & Latorre, 1978; El Hady & Machta, 2015; P.-C. Zhang et al., 2001) of the neural activity. We apply mechanical loads inducing damage (Cinelli, Destrade, Mchugh, et al., 2017; Jérusalem et al., 2014) at the nerve membrane layer to investigate the changes in neuronal membrane excitability (Jérusalem et al., 2014) and propagation (Boucher et al., 2012) in response to changes in electrostriction (Mueller & Tyler, 2014). The electrical and the mechanical fields of the model are coupled by using electro-thermal equivalences and equivalent materials properties in FE analysis (Cinelli, Destrade, Duffy, et al., 2017b; Cinelli, Destrade, Mchugh, et al., 2017).

We achieve coupling of the electro-mechanical effects of the action potential (Hodgkin & Huxley, 1952) by modelling the nerve membrane as a piezoelectric material (P.-C. Zhang et al., 2001), and implementing the thermal analogy of the neural activity (Cinelli, Destrade, Duffy, et al., 2017b; Cinelli, Destrade, Mchugh, et al., 2017).

In contrast to (Cinelli, Destrade, Mchugh, et al., 2017), here we analyse the effects of nerve bundle geometry and type on the electro-mechanical coupling to evaluate permanent electro-mechanical impairments due to plasticity when mechanical loads are applied. Although nerve axons show some elastic recovery of the pre-stretched geometry under slow loading rates (Smith et al., 1999), the role of plasticity in delaying the mechanical response is fundamental in understanding the pathology due to stretch injury occurring at fast loading rates. Indeed, induced-permanent focal axonal dysfunction and induced-permanent focal electrophysiological impairments may explain the adaptive recovery of neural connections seen in mild-to-moderate TBI, and the potential synaptic rearrangements seen in severe TBI (Jafari, Nielson, David, & Maxwell, 1998; H. C. Wang & Ma, 2010; J. Wang, Hamm, & Povlishock, 2011).

We also show that variability in axonal calibre affects axonal vulnerability, leading to differential injury responses in myelinated and unmyelinated axons (Hemphill et al., 2015; Perge, Niven, Mugnaini, Balasubramanian, & Sterling, 2012). Morphological changes of the cellular structures are more likely to happen in unmyelinated than myelinated axons (Hemphill et al., 2015; Jafari et al., 1998; Reeves, Smith, Williamson, & Phillips, 2012), and they occur in the form of molecular-based processes such as leaking nerve membrane (Yu et al., 2012) and cytoskeleton disruption (Hemphill et al., 2015; Jafari et al., 1998; Smith et al., 1999; Tang-schomer, Patel, Baas, & Smith, 2017). Unmyelinated axons are at greater risk compared to myelinated axons, where injuries occur preferentially at the Ranvier node regions (Hemphill et al., 2015). Additionally, larger calibre axons are shown to be more vulnerable to injury due to their higher metabolic requirements, and they are more prone to develop pathologies (Hemphill et al., 2015; Reeves et al., 2012).

Our proposed 3D finite element model of a nerve bundle includes a representation of a nervous cell made of extracellular media (ECM), a membrane, and intracellular media (ICM). The bundle model is a section of an idealized geometry of a nerve bundle consisting of four identical parallel cylindrical unmyelinated or myelinated fibres, see Figure 1. The diameters of these fibres are within the range of the human optic axon (Perge et al., 2012). We consider the case of two scaled nerve bundle models with a ratio of 2:1, where the nerve fibres inside follow the same ratio, keeping the same thickness for the nerve membrane (Cinelli, Destrade, Duffy, et al., 2017b; Cinelli, Destrade, Mchugh, et al., 2017). The bundles are made of identical unmyelinated or myelinated nerve fibres. We use different sizes and nerve types to enhance the understanding of neurotrauma in mechanically-injured bundles, as revealed in experiments at the cellular level (Galbraith et al., 1993; Geddes et al., 2003; Yu et al., 2012).

The use of a 3D geometry with plastic material properties allows for the simulation and evaluation of strain and voltage distributions before and after the induced damage. The inclusion of 3D mechanically-induced electrophysiological impairments is needed to enhance the understanding of electro-mechanical changes in neurotrauma evaluation, and improve diagnosis, clinical treatment and prognosis (Lajtha, 2009; Ma et al., 2016). This approach might prove crucial to study and understand the mechanics at play in neuro-physiology, as observed experimentally in damaged nerve membranes of clinical cases such as multiple sclerosis (Demerens et al., 1996; Galbraith et al., 1993; Geddes et al., 2003).

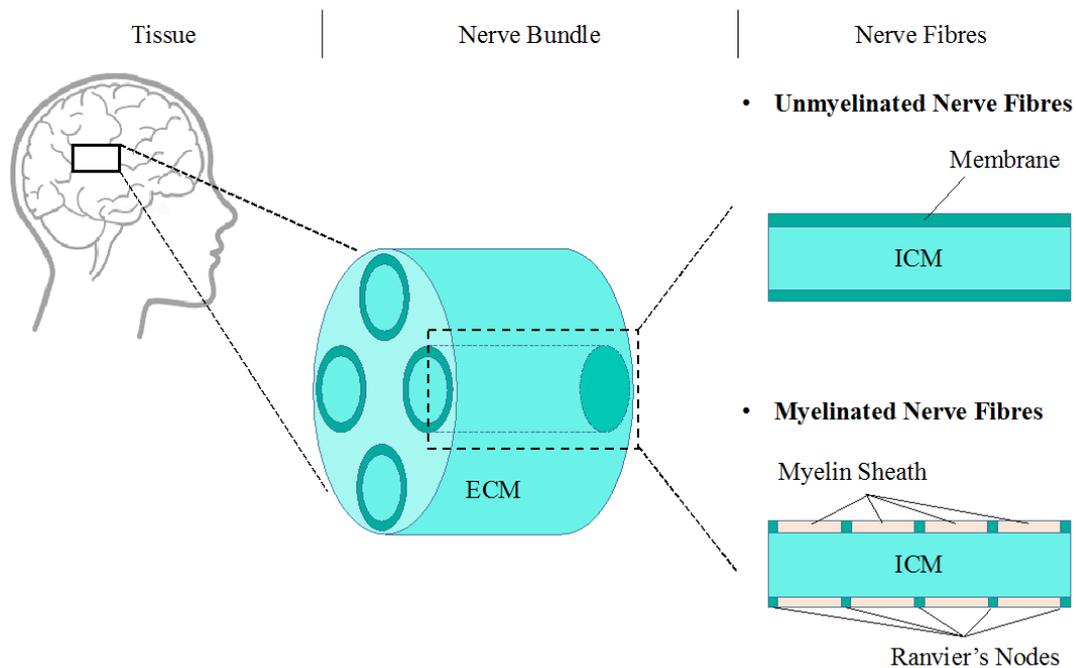

Figure 1: Sketch of the nerve bundle geometry. The bundle is made of four identical fibres (ECM = extracellular media, ICM = intracellular media).

## II. Methods

### A. Model

With the purpose of highlighting the importance of a unified electro-mechanical theory for neural applications (El Hady & Machta, 2015; Heimburg, Blicher, Mosgaard, & Zecchi, 2014; Hemphill et al., 2015; Mosgaard, Zecchi, & Heimburg, 2015; Mueller & Tyler, 2014), a modelling approach (Cinelli, Destrade, Duffy, et al., 2017b; Cinelli, Destrade, Mchugh, et al., 2017) is further developed in this paper to replicate electro-mechanical phenomena accompanying the neural electrical activity (Cinelli, Destrade, Duffy, et al., 2017b; Cinelli, Destrade, Mchugh, et al., 2017; El Hady & Machta, 2015; Mueller & Tyler, 2014). Here, the use of an idealized geometry of a nerve bundle is meant to reduce complexity and computational cost arising from the use of 3D morphological images of neuronal structure (Cinelli, Destrade, Duffy, et al., 2017b; Lytton et al., 2017).

Our bundle model simulates the exchange of charges in four identical cylindrical neurites, made of an intracellular media (ICM) enclosed by a thin membrane, and surrounded by extracellular media (ECM) (Cinelli, Destrade, Duffy, et al., 2017b; Cinelli, Destrade, Mchugh, et al., 2017), as shown in (Cinelli, Destrade, Mchugh, et al., 2017). Additionally, two fibre bundle models are considered, scaled in size in the ratio of 2:1. Here, only the cases of a fully unmyelinated bundle or a fully myelinated bundle are considered, but the same process may be applied to investigate mixed fibre bundles.

Regardless of the diameter, the nerve membrane layer has a finite thickness equal to 3 nm, representing the dielettric part of the membrane only (Belanger & Thornton, 2013; Cinelli,

Destrade, Duffy, et al., 2017b). Instead, the myelin layer of myelinated fibre is a periodically-partitioned region at the nerve membrane layer, with the same finite thickness (Cinelli, Destrade, Mchugh, et al., 2017; Einziger, Livshitz, Dolgin, & Mizrahi, 2005), see Figure 1. The reason for this assumption is motivated by the use of the Cable Equation for simulating diffusion of charges in heterogeneous conductors as discussed by Einzinget et al. (Einziger et al., 2005). Thus, the Cable Equation models charge diffusion in homogeneous and heterogeneous conductors (i.e. unmyelinated and myelinated fibres, respectively), and this can be implemented as an equivalent thermal process in finite element analysis (Cinelli, Destrade, Duffy, et al., 2017b; Cinelli, Destrade, Mchugh, et al., 2017).

B. Material Properties

We assume incompressible isotropic mechanical properties (El Hady & Machta, 2015). We also assume the same isotropic plastic behaviour for the nerve membrane, ICM, and myelin layer. The yield stress is calculated with an engineering strain equal to 21 % (Allison C Bain & Meaney, 2000) and a Young Modulus equal to 1GPa (El Hady & Machta, 2015). Strain hardening is assumed to occur up to a strain of 65 % (Smith et al., 1999). Thus, the engineering strain and engineering stress values are $(0.21, 0.21 \text{ GPa})$ and $(0.65, 0.65 \text{ GPa})$ for the yield strain limit and strain hardening, respectively. Beyond 65 % strain, the stresses are assumed to remain constant.

The electrical model parameters for unmyelinated and myelinated fibres are taken from (Cinelli, Destrade, Mchugh, et al., 2017) and (Jérusalem et al., 2014), respectively. This model assumes that the exchange of charges occurs in the through-thickness direction of the nerve membrane, rather than along the fibre length (Cinelli, Destrade, Mchugh, et al., 2017; Hodgkin & Huxley, 1952; P.-C. Zhang et al., 2001). So, the piezoelectric effect is only relevant in the through-thickness direction, represented here with orthotropic piezoelectric constants of approximately 1 nm per 100 mV (P.-C. Zhang et al., 2001) in the thickness direction and zero in the longitudinal and circumferential directions, while the electrical capacitance per unit area, $C_m$, changes as the square of the voltage (Alvarez & Latorre, 1978; Cinelli, Destrade, Duffy, et al., 2017b; Cinelli, Destrade, Mchugh, et al., 2017; El Hady & Machta, 2015).

C. Implementation

We expand on the Hodgkin and Huxley (HH) model to include 3D fields, elasticity and plasticity, see contrast depicted in Figure 3. With the electro-thermal equivalences (Cinelli, Destrade, Duffy, et al., 2017b; Cinelli, Destrade, Mchugh, et al., 2017), we can visualise in 3D the neural activity, the distribution of voltage and the generated strain, using well-established coupled thermo-mechanical software simulation tools (Cinelli, Destrade, Duffy, et al., 2017b; Cinelli, Destrade, Mchugh, et al., 2017). This model is implemented as a coupled thermo-mechanical model in the finite element software code Abaqus CAE 6.13-3, where electricity is simulated as thermal analogy (Cinelli, Destrade, Duffy, et al., 2017b).

Then, by using user-defined subroutines (Cinelli, Destrade, Duffy, et al., 2017b; Cinelli, Destrade, Mchugh, et al., 2017), thermal equivalent electrical properties are assigned to the membrane of each fibre in the bundle, independently, based on the spike initiation (Platkiewicz & Brette, 2010), strain (Boucher et al., 2012; Jérusalem et al., 2014) and voltage (Alvarez & Latorre, 1978) generated at each membrane. As in (Cinelli, Destrade, Mchugh, et al., 2017), the membrane neural activity changes in response to the membrane voltage V, total strain, ε, at the membrane (Hodgkin & Huxley, 1952; Jérusalem et al., 2014), space and time. In contrast to (Cinelli, Destrade,

Mchugh, et al., 2017), here, total strain includes elastic, piezoelectric (thermal equivalent (Cinelli, Destrade, Duffy, et al., 2017b)) and plastic strain, see Figure 3.

The HH reversal voltage potentials of sodium, $E_K$, and potassium, $E_{Na}$, change due to voltage and strain at the nerve membrane (Boucher et al., 2012; Jérusalem et al., 2014), and hence the threshold of action potential initiation changes (Hodgkin & Huxley, 1952). In particular, the axial component of the total strain, read along the fibre length, links mechanical loads and electrical activity in nervous cells (Cinelli, Destrade, Mchugh, et al., 2017; Jérusalem et al., 2014) inducing changes in reversal potentials at the nerve membrane, as discussed by Jérusalem et al. (Jérusalem et al., 2014). Then, the reversal potential of the leak ions $E_{l^-}$ is not influenced by the strain but varies based on changes in the gradient concentrations of potassium and sodium across the membrane (Jérusalem et al., 2014).

For traumatized channels, the changes in conductivity for sodium, $G_K$, and potassium ions, $G_{Na}$, follow the changes in the respective reversal potentials (Hodgkin & Huxley, 1952). Additionally, the nerve membrane integrity varies with the fraction of nodal channels (AC) affected by the trauma, while the other membrane's channels, $(1 - AC)$, remain intact (Boucher et al., 2012), see (Cinelli, Destrade, Duffy, et al., 2017b; Cinelli, Destrade, Mchugh, et al., 2017). Here, only the extreme cases of the entire membrane being traumatized (AC = 1) or intact (AC = 0) are shown as illustrative examples.

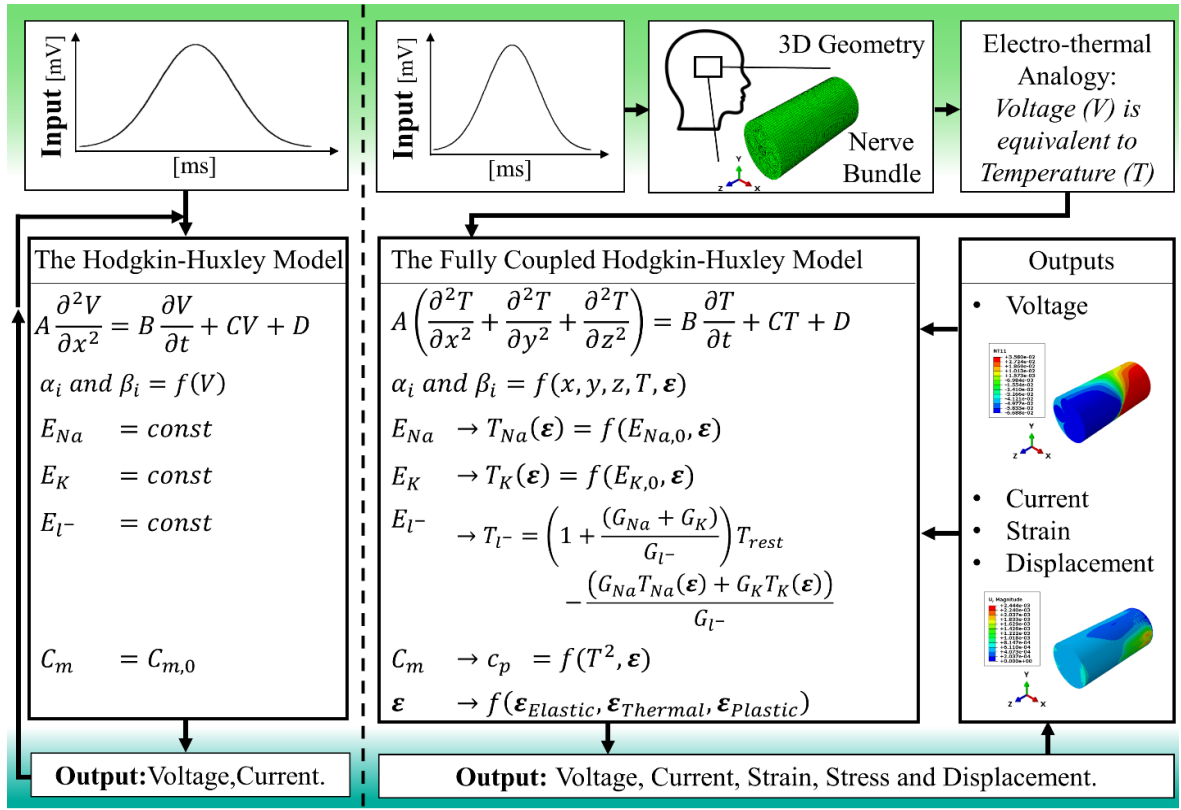

*Figure 2: Flowchart of the code describing the active behaviour of the nerve's membrane: on the left, the uncoupled Hodgkin and Huxley (HH) dynamics* (Hodgkin & Huxley, 1952) *and on the right, the fully coupled HH dynamics used in this paper. A Gaussian voltage distribution elicits*

*the action potential in a 3D model of a nervous bundle. By electro-thermal equivalence, the full HH model is implemented as an equivalent thermal process, with changes in the membrane's conductivity* (Hodgkin & Huxley, 1952)*, the capacitance* (Cinelli, Destrade, Duffy, et al., 2017b) *and the HH parameters* (Jérusalem et al., 2014)*.*

### D. Boundary Conditions

For evaluating the influence of neuron structure in neurotrauma, we assume only one active fibre in each bundle, Fibre#3, while the other fibres are activated by the diffusion of charges when the induced damage is minimal or absent (Cinelli, Destrade, Mchugh, et al., 2017). For the purposes of this paper, the one-fibre activation assumption allows for quantifying the generated electro-mechanical alterations (Cinelli, Destrade, Mchugh, et al., 2017) in a simplified context, as first step towards a more detailed analysis. This assumption allows for understanding the distribution of charges when mechanical damage is induced, focusing on the radial distribution of charges rather than on longitudinal.

In particular, an upper-threshold stimulation voltage with a Gaussian distribution (Cinelli, Destrade, Mchugh, et al., 2017; Tahayori, Meffin, Dokos, Burkitt, & Grayden, 2012) is applied on Fibre #3 along its length, see Figure 2, while the other fibres are activated only if the diffused charges from Fibre #3 generate an input voltage higher than the modulated threshold (Platkiewicz & Brette, 2010). The 3D distribution of charges on Fibre #3 modulates the activation of the other fibres.

We consider in turn two cases of applied mechanical loads at the bundle. As a first step to assess the inclusion of plasticity using this novel coupling method, only frequency-independent loading conditions are considered throughout, following the initial steady-state regime (lasting about 2ms) (Cinelli, Destrade, Duffy, et al., 2017b; Cinelli, Destrade, Mchugh, et al., 2017). The mechanical loads are applied from 2ms to 67 ms, as instantaneous loading conditions, and the model runs for 140 ms so that the effects of plasticity can be observed post-loading. An encastré boundary condition is enforced at the origin of each model, so that no movement and rotation is allowed at the origin node. Then, no rotation is allowed for all the nodes at the origin bundle side.

In the first case, we apply an instantaneous uniform compression to the bundle to simulate injury conditions, with two values of pressure, simulating mild (less than 55 kPa) and severe (higher than 95 kPa) pressures (Hosmane et al., 2011).

In the second case, we reproduce the axial strain conditions of the uniaxial test conducted by Bain and Meaney (Allison C Bain & Meaney, 2000). Two values of instantaneous uniform stretch are applied as a displacement boundary condition to simulate 5 % and 14 % of total axial deformation, ε, where the probability of inducing morphological injury during the elongation test is 5% and 25%, respectively (Allison C Bain & Meaney, 2000). Additionally, we also consider the cases of 25%, 30% and 60% elongation to investigate the electro-mechanical responses within the range of plasticity (before mechanical failure) (Smith et al., 1999).

### III. Results

#### I. Pressure Loads

Figure 3 (a) shows the membrane potential of a small bundle made of unmyelinated nerve fibres (SBUN), and Figures 3 (c)-(e) show the radial displacement in small (SBUN) and big (BBUN) unmyelinated bundles under mild (25kPa) and severe (192kPa) pressures inducing axonal injuries (Hosmane et al., 2011). Figure 3 (b) shows the membrane potential of a small myelinated bundle (SBMY). Figures 3 (d)-(f) show the radial displacement in small (SBMY) and big (BBMY) myelinated bundles, for each pressure case. Note that we find the same voltage responses for both small and big bundle models, because they have the same membrane properties. Here, the strain applied at the nerve membrane by compressing the bundle shifts the ionic reversal potentials of the fully coupled HH model by a quantity which varies depending on the magnitude of the applied load, see flowchart in Figure 2. Results are taken at the maximum radial displacement on Fibre #3.

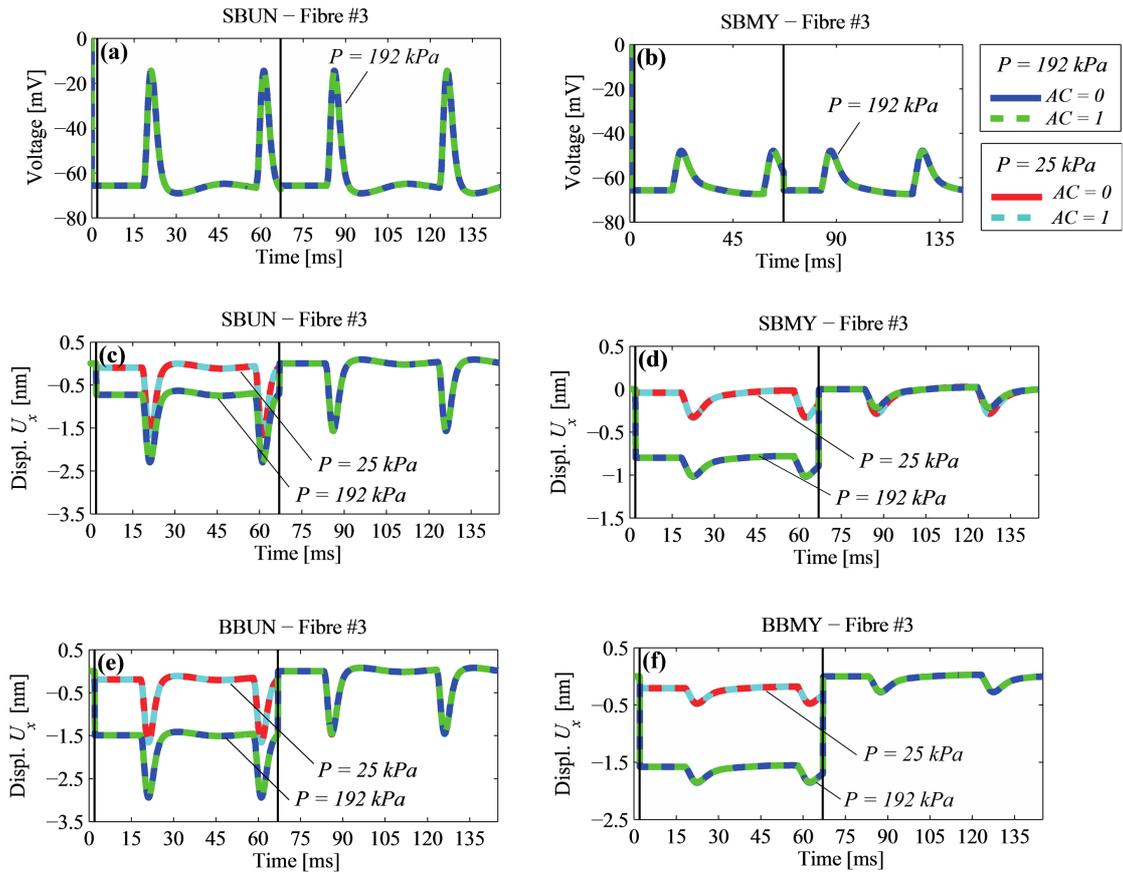

*Figure 3. (a) Membrane Potential [mV] on Fibre #3 in a small unmyelinated bundle (SBUN). (b) Membrane Potential [mV] on Fibre #3 in a small myelinated bundle (SBMY). Radial displacement [nm], $U_x$, of unmyelinated small (c) and big (e) bundles, myelinated small (d) and big (f) bundles. The uniform applied loads are classified as mild (25kPa) and severe (192kPa) pressures* (Hosmane et al., 2011). *Data are the maximum radial displacement of a node on Fibre #3 in both cases.*

We find that mild (25 kPa) and severe (192 kPa) pressure levels have a similar impact on the signal transmission, both in terms of reduced magnitude and shift over time, due to similar strain values read at the nerve membrane. In both small (SBUN) and big (BBUN) unmyelinated bundles, the action potential peak is $-14.4$ mV at 19 ms, see Figure 3 (a), while in small (SBMY) and big (BBMY) myelinated bundles, the peak is $-47.54$ mV at 22 ms, see Figure 3 (b).

For any applied load, the reversal voltage potentials are changed due to the induced-strain in the bundle and the magnitude of the action potential is reduced (Jérusalem et al., 2014). The applied compression leads to changes in reversal potentials, according to the strain values read along the bundle middle axis only (Jérusalem et al., 2014). In all cases, only slight differences are found for a traumatized ($AC = 1$) compared to a non-traumatized nerve membrane ($AC = 0$) when mild or severe pressures are applied, see Figures 3 (a)-(b). The $AC$ variable impacts the ionic conductance (Boucher et al., 2012) whose changes are not contributing to the total strain along the bundle middle axis direction during compression. This is because we modelled the piezoelectricity of the membrane radially rather than longitudinally (Cinelli, Destrade, Duffy, et al., 2017b; P.-C. Zhang et al., 2001), as discussed in (El Hady & Machta, 2015; Hodgkin & Huxley, 1952; P.-C. Zhang et al., 2001). Indeed, the applied pressures lead to an axial displacement of less than 1 % of the total length of the bundles in each model. Despite the size of the bundle, traumatized nerve membranes seem to be able to carry and generate signals both during and after uniform compression.

Figures 3 (c) and (e) show the radial displacement on Fibre #3 in SBUN and BBUN, respectively. While the shift in baseline displacement is proportional to the applied pressure in both cases, the amplitude of the peak value from the baseline follows the membrane voltage response, and therefore values for the unmyelinated bundles are the same regardless of the bundle size, see Figures 3 (c) and (e).

In Table 1 we collected the values of the displacement shift, during and after loading, at mild (25 kPa) and severe (192 kPa) pressure levels. We also computed the maximum values of the plastic strain ($PE$) and total strain ($E_{tot}$) once the loads are removed.

| Boundary Conditions | Intensity | Unmyelinated Nerve Bundles | | Myelinated Nerve Bundles | |
|---|---|---|---|---|---|
| | | SBUN | BBUN | SBMY | BBMY |
| | | **Displacement Peak [nm]** | | | |
| During Loading | 25 kPa | $-1.67$ | $-1.65$ | $-0.34$ | $-0.48$ |
| | 192 kPa | $-2.30$ | $-2.95$ | $-1.02$ | $-1.85$ |
| After Loading | 25 kPa | $-1.47$ | $-1.47$ | 0 | 0 |
| | 192 kPa | $-1.47$ | $-1.47$ | $-0.28$ | $-0.26$ |
| | | **Strain [%]** | | | |
| Plastic Strain | 25 kPa | 0.5 | 0.5 | 0 | 0 |
| | 192 kPa | | | | |
| Max. Total strain | 25 kPa | 3.5 | 3.5 | 1.5 | 1.5 |
| | 192 kPa | | | | |

*Table 1. Displacements and strains on Fibre#3 when compression is applied.*

Figures 4 (a)-(d) show the difference in voltage distribution over the bundle, at the action potential peak, in small and big, myelinated and unmyelinated bundles, with non-traumatised nerve membranes ($AC = 0$) bundles, under mild pressure (25 kPa).

The myelinated bundles (Figures 4 (a) and (c)) experience uniform compression. At the peak of the membrane potential applied on Fibre #3, the piezoelectric effect generates an additional contraction on its nerve membrane, dragging parts of Fibres #2 and #4 in its vicinity. Thus, we find four peaks of the maximum total strain $E_{tot}$ on Fibre#3: two are at the regions in proximity to Fibres #2 and #4; and two in the diametrically opposed regions (to conserve the overall volume by incompressibility). Then, on the encastré side of the bundle, where the applied voltage is higher, local regions of high voltage are found on Fibres #2 and #4, due to their vicinity with Fibre #3. So, local contractions on Fibres #2 and #4 act in opposition to the contractions on Fibre#3. The voltage distribution in Figures 4 (a) and (c) is influenced by the final distribution of $E_{tot}$. Hence, the voltage is higher at the regions where: the applied voltage Gaussian distribution is high, the nerve membrane is not constrained and the strains are low.

A similar scenario occurs for myelinated bundles, see Figures 4 (b) and (d). However, here the piezoelectricity is limited to the Ranvier node regions, whose displacement is constrained by the myelin layer. Accordingly, the $E_{tot}$ is more uniform at the nerve membrane layer of all the fibres than for unmyelinated bundles. Fibres surrounding the active Fibre #3 are not activated because the charge read at their nerve membrane is lower than the minimum threshold for activation (Hodgkin & Huxley, 1952).

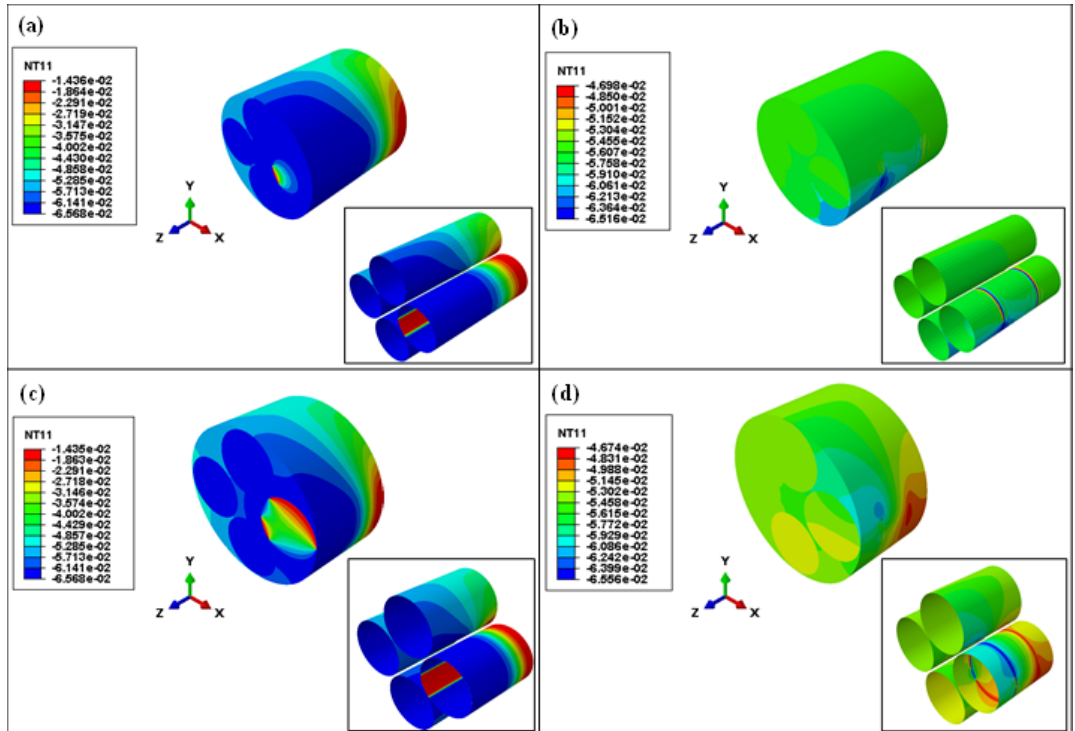

*Figure 4: Voltage distribution (NT11), at the action potential peak, in small (a) and big (c) myelinated bundles, and small (c) and big (d) myelinated bundles for 25kPa pressure with AC=0 (intact membranes). The small box shows the membrane layer of each model.*

II. Displacement Loads

Figures 5-8 illustrate the response of a bundle to a displacement boundary condition applied at one end along its fibre length, simulating 5 %, 14 %, 25 %, 30 % and 60 % elongation of the total length of the bundle (in line with the experiments conducted by Bain and Meaney (Allison C Bain & Meaney, 2000)). Those elongation values are chosen to initiate an elastic response if the applied strain is lower than 21 % (Allison C Bain & Meaney, 2000), or a plastic response if it is higher (according to the experiments conducted by Smith et al.(Smith et al., 1999) on cultured axons). The Figures show the membrane potential at the position of maximum displacement on Fibre #3, the maximum principal value of plastic strains ($PE_z$) along the bundle middle axis (Jérusalem et al., 2014) and the total strain $E_{tot}$.

Figure 5 shows the membrane voltage distribution in small (a) unmyelinated (SBUN) and (b) myelinated bundles (SBMY). In all elongations, we find that the action potential signal is reduced significantly if not eliminated altogether, while there is a clear increasing shift in baseline voltage with increasing applied displacement. The effect of strain on the baseline shift is higher for the myelinated than for the unmyelinated bundles. This finding is in line with previous results, where the voltage read at the nerve membrane varies linearly with the elastic component of the total strain up to 21 % (Allison C Bain & Meaney, 2000) when permanent electrophysiological impairments (such as leaking ionic channels (Boucher et al., 2012; Yu et al., 2012)) alter the osmotic gradient across the membrane, and so the ability to carry and generate action potentials (Galbraith et al., 1993; Jérusalem et al., 2014; Smith et al., 1999). Then, the reversal potentials change depending on the level of elongation, the type of bundle and its size, due to the different distribution of total strain within the bundle.

Additionally, the fraction of nodal channels affected by trauma $AC$ induces differences in the membrane voltage peaks from the membrane baseline, see Figures 5 (a) and (b), while there are little or no differences membrane baseline value with varying $AC$. In the SBUN case, increasing the fraction $AC$ increases the difference between the membrane potential peak and the membrane baseline at higher strains only in small unmyelinated bundles. For example, at 60 %, the membrane potential is $-24.15\ mV$ with $AC = 0$ and $-2.134\ mV$ with $AC = 1$, see Figure 5 (a). In small myelinated bundles (SBMY), instead, the membrane peaks at 25 % are about the same as the values found at 30 and 60 % of elongation, see Figure 5 (b). Similarly to SBUN, the maximum voltage in SBMY at 60 % elongation is $-23.21\ mV$ with $AC = 0$ and $-2.08\ mV$ with $AC = 1$, see Figure 5 (b). Then, at high applied strain, the voltage peaks in big unmyelinated bundles (BBUN) show similar differences when going from $AC = 0$ to $AC = 1$ ($-24.17\ mV$ with $AC = 0$ and $-2.15\ mV$ with $AC = 1$, not shown here), while in big unmyelinated bundles (BBMY), the peaks are about the same order regardless of $AC$ ($-4\ mV$ with $AC = 0$ and $-1\ mV$ with $AC = 1$, not shown here). However, big bundles show higher shift in membrane potential at 30 % elongation, which is the maximum elongation for big bundles in this study.

These trends are to be expected, because the fraction of nodal channels affected by trauma $AC$ impacts the ionic conductance of the nerve membrane (Boucher et al., 2012). Hence, with $AC = 1$, a membrane plateau is reached because of the combined changes in ionic conductance (Boucher et al., 2012) and in reversal potentials (Jérusalem et al., 2014). The plateau shape occurs due to the similar values of the reversal potentials in a membrane in which the ion mobility of potassium and sodium are similar.

In unmyelinated fibres, the membrane voltage is shifted following elongation and no action potential is generated because of the higher strain components along the fibre length, where the strain is highest (Jérusalem et al., 2014). In contrast, similarly to compression, the myelin layer induces a different distribution of strain at the Ranvier node regions of the nerve membrane layer, where strains along the bundle middle axis are lower than in unmyelinated bundles, as observed in experiments (Hemphill et al., 2015; Jafari et al., 1998; Reeves et al., 2012). Therefore, those fibres are more likely to generate an action potential after elongation. This is the reason why in SBMY, the potential does not have a plateau shape when $AC = 1$ as in myelinated fibres, see Figure 5. Then, in the BBMY, the fibre is still able to generate action potential, both during and after loading, although there is a shift in membrane voltage baseline, as seen in the other cases. This suggests that the myelin layer induces a different distribution of strain within the fibre to preserve its functionality.

After the load is removed (i.e. after $67\ ms$), only in the case of 60 % elongation is the membrane baseline for SBUN shifted, up to $-45.7\ mV$ ($AC = 0$) and to $-41.7\ mV$ ($AC = 1$), as the ionic gate channels are kept open by the permanent plastic strains at the nerve membrane. Similarly for SBMY after loading, the membrane voltage baseline goes to $-55.64\ mV$ only for 60 % elongation ($AC = 0$ and $AC = 1$), while it remains about $-65\ mV$ for the other elongation values, see Figure 5. As seen by Jérusalem et al. (Jérusalem et al., 2014), we find here that the larger the elongation, the greater the shift of the membrane potential to a higher peak value, where the reversal potentials are affected differently by the strain magnitude.

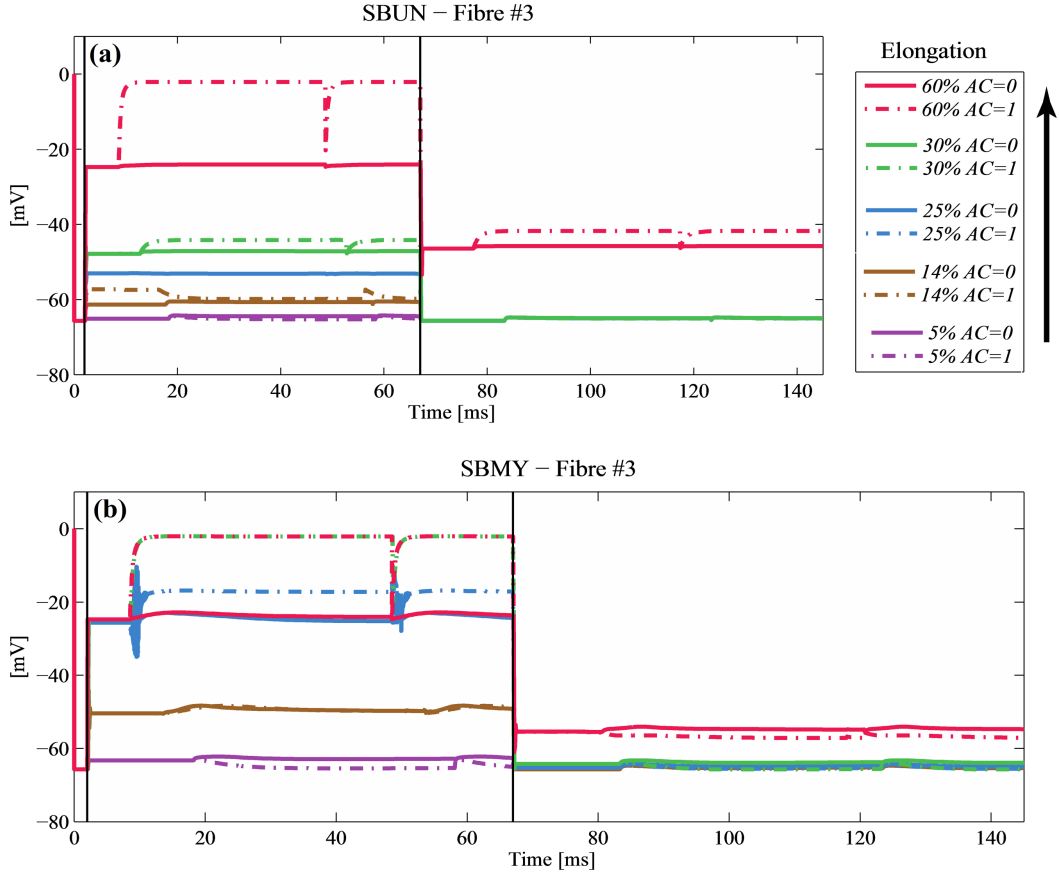

*Figure 5. Membrane Potential [mV] on Fibre #3 in small (a) unmyelinated (SBUN) and (b) myelinated (SBMY) bundles, under 5% to 60% elongation. The load is applied from time t=2 ms to time t=67 ms. The fraction of ionic channels affected by the strain is AC=0 is for an intact membrane and AC=1 for a traumatized membrane* (Boucher et al., 2012). *Data are taken at the maximum displacement along the bundle middle axis on Fibre #3.*

This effect can be better appreciated in Figures 6 and 7, where membrane voltage peaks and corresponding strains are taken at the node of maximum axial displacement and the node of maximum peak voltage on Fibre#3, respectively. Figure 6 shows the peak of the membrane potential in (a) small and (c) big bundles, and the corresponding plastic strains, (b) and (d), during elongation, at the node of maximum axial displacement along the bundle middle axis (i.e. along the $z-axis$). The membrane potential, see Figures 6 (a)-(c), shows only slight changes with the fraction of nodal channels affected by trauma $AC$ for strains lower than 30 %. Although the voltage in SBMY levels off at lower strains than in SBUN, at 60%, the differences in membrane potential are not dependent on fibre type, but only on $AC$, see Figure 6 (a). In SBUN and SBMY, the same value of maximum potential is reached at 60 %, which is about $-20\ mV$ with $AC = 0$ and about $0\ mV$ with $AC = 1$, see Figure 6 (a). The corresponding $PEs$ are lower than 25 % for applied elongation values lower than 30 %, while they are much higher in SBUN (143 %) than in SBMY (20 %) at 60 % elongation, see Figure 6 (b). However, the local maximum peak voltage, relative to Fibre#3, reaches a constant value above 25 % elongation, see Figure 7 (a). Differences in maximum at lower elongations, see Figure 8 (a), are due to elastic and thermal strains (i.e. the thermal equivalent of piezoelectric strain (Cinelli, Destrade, Mchugh, et al., 2017)), see Figure 7 (b), because of the small plastic strains found for the same conditions, see Figure 6 (b).

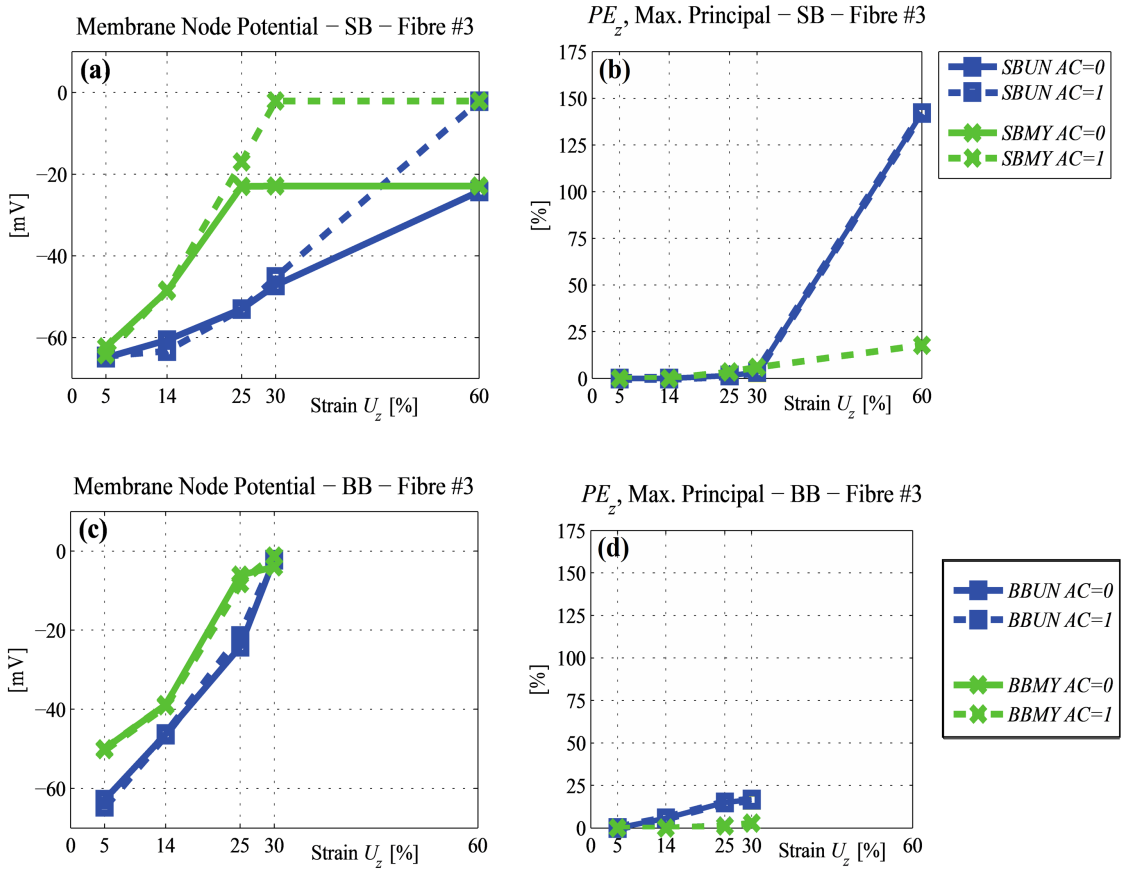

*Figure 6. Peak values of the membrane potential [mV] in (a) small and (c) big bundles; and maximum values of the maximum principal plastic strain at the node of interest, along the bundle axis, for (b) small and (d) big bundles. Data are taken on Fibre #3 at the node where the displacement along the bundle axis is maximum. The axial elongations are 5%, 14%, 25%, 30% and 60%. In big bundles, failure occurs at 30% applied displacement.*

In contrast, when doubling the size of the bundles, the changes in membrane potential are of the same order of magnitude for unmyelinated and myelinated bundles in all elongation cases. Additionally, in contrast to what happens to small bundles, $AC$ has a little influence on these voltage variations, see Figure 6 (c).

Similarly, the local maximum peak voltage read on Fibre#3 does not show great variation with either fibre type or $AC$ during elongation tests, see Figure 7 (a). Again, slight differences in plastic strains ($PE$) are found going from $AC = 0$ to $AC = 1$, see Figures 6 (b) and (d), suggesting that the changes in ionic conductance (Boucher et al., 2012) has a smaller impact, compared with the applied strain, in generating plastic strain at the nerve membrane. At high strain (here, at 30 %), the peaks, read in both bundles, reach the same value, independent of the fibre type and $AC$ value, see Figure 6 (c). Although there is a large shift in potential, the plastic strains are lower than 25 % in both BBUN and BBMY. At the same applied strains, the plastic strain is not the main component of the maximum value of the total strain $E_{tot}$ read at the bundles, see Figure 7 (d), which are up to 75 % in BBUN and 120 % in BBMY.

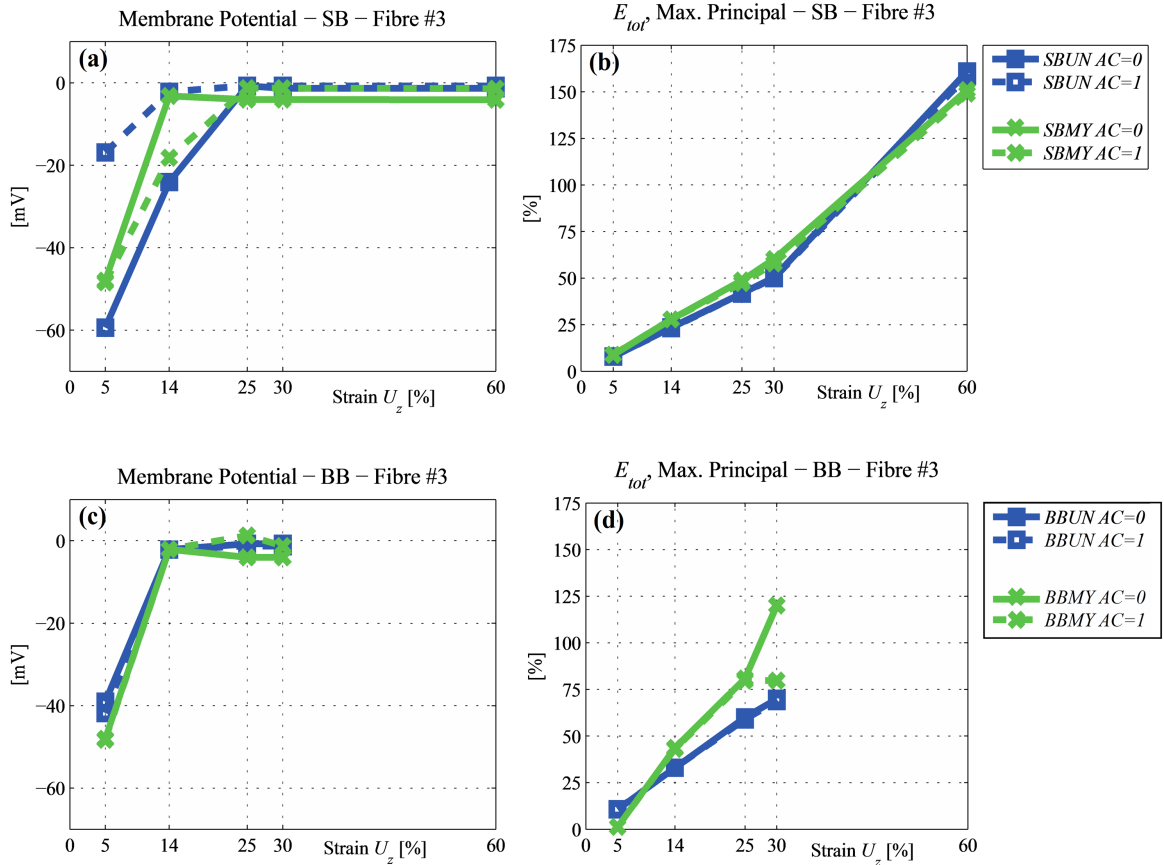

*Figure 7. (a),(c) show the peak values of the membrane potential [mV] found on Fibre#3, in the SB and BB, respectively. In (b),(d), the maximum value of the total strain ($E_{tot}$) is the maximum found on Fibre#3. Data are taken at the maximum peak voltage in all the cases. Strains are 5%, 14%, 25%, 30% and 60%. In BB, failure occurs at 30% applied displacement.*

Data show that plastic strains are responsible for the functional and mechanical failure in small unmyelinated bundles (SBUN), while functional recovery is more likely to happen in small myelinated bundles (SBMY) because the elastic component is 70 % of the $E_{tot}$ at 60 % elongation, see Figure 7 (b). Then, larger bundles show higher variation in membrane voltage due to the higher number of changes exchanged per unit area on the fibre. However, the $E_{tot}$ at 30 % elongation are comparable to the values found in small bundles at 60 %, despite the lower plastic strains (*PE*) at the membrane. Failure occurs for elongation higher than 30 % suggesting that the $E_{tot}$ are mainly along the bundle middle axis. Strains along the fibre length are those with greater impact on the membrane voltage (Jérusalem et al., 2014), and are responsible for the voltage shift in all the bundles considered here.

Then, myelinated bundles (both SBMY and BBMY) show smaller $PE$ in contrast with unmyelinated bundles (both SBUN and BBUN) at the same elongation. Particularly, at 30 % elongation, the plastic strain is less than 10 % in myelinated bundles (5.90 % in SBMY and 2.88 % in BBMY) and greater than 15 % in unmyelinated bundles (15.42 % in SBUN and 16.56 % in BBUN). The myelin layer seems to redistribute the induced permanent damage on the whole fibre, rather than on the fibre length, as in unmyelinated bundles. Thus, at the same loading conditions, myelinated fibres and bundles are stronger than the unmyelinated ones.

Figures 8 (a)-(d) show the difference in total displacement distribution, at the action potential's peak for 30% elongation and no traumatized nodal channels ($AC = 0$). Similarly to Figure 4, the voltage distribution is affected by the total generated strain $E_{tot}$ at the nerve membrane layer of each fibre. Here, the fibres are pulled along the bundle middle axis while Fibre#3 is contracting, dragging Fibre#2 and #4. The maximum value of the total strain $E_{tot}$ is lower around the centre of the bundle due to the balance of negative radial and positive axial strains. Here the plastic strains ($PE$) are consistently higher than those generated by applied pressure, see Figure 3. Again, the voltage is higher in the regions where lower strains are found.

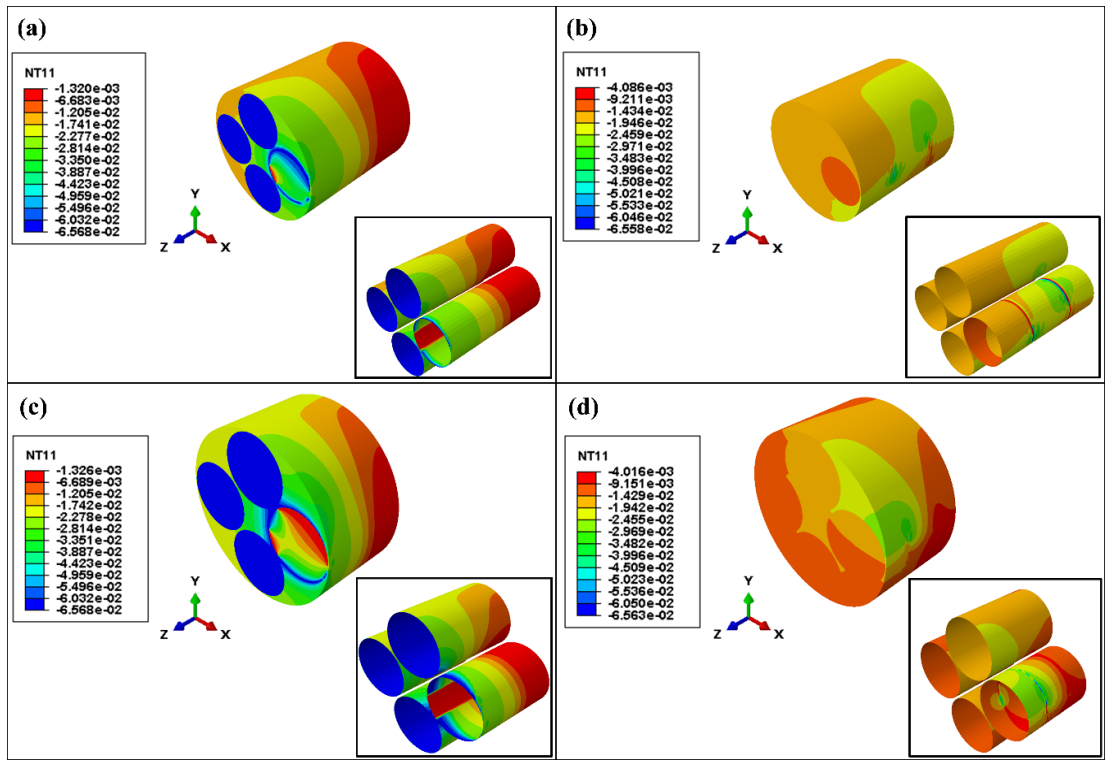

*Figure 8. Isometric view of (a) SBUN and (b) SBMY for 60% elongation; and, (c) BBUN and (d) BBMY for 30% elongation with AC=0. Data are taken at the peak of the action potential. NT11 is the equivalent voltage variable [in V]. The box shows the membrane layer of each model.*

## III. DISCUSSION

Beginning with the original Hodgkin and Huxley (HH) study in 1952, numerous studies have been modelled neural activity as a purely electrical phenomenon. The inclusion of the electro-mechanical coupling (Cinelli, Destrade, Duffy, et al., 2017b; Cinelli, Destrade, Mchugh, et al., 2017; El Hady & Machta, 2015) (such as electrostriction (Alvarez & Latorre, 1978) and piezoelectricity (P.-C. Zhang et al., 2001)) accompanying neural activity aims at improving the understanding of neuron-to-neuron communications, diseases and clinical treatments (Ma et al., 2016; Shi & Pryor, 2002; Wright & Ramesh, 2012). Computational modelling is a powerful research tool to investigate and simulate these complex phenomena.

In contrast to previous studies (Jérusalem et al., 2014; Wright & Ramesh, 2012), this paper shows the advantages of a fully coupled electro-mechanical 3D framework to investigate the details of neural activity, combining real-time fully coupled electro-mechanical phenomena, a modulated threshold for spiking activation and independent alteration of the electrical properties for each fibre in the 3-layer nerve bundle (Cinelli, Destrade, Duffy, et al., 2017b; Cinelli, Destrade, Mchugh, et al., 2017). The use of coupled electro-mechanical finite element modelling for neural engineering (Cinelli, Destrade, Duffy, et al., 2017b) opens the way to a different investigation of the neuron nature itself. The use of a 3D geometry allows for a physical representation of the neuron cell and morphology in signal propagation with trauma (Cinelli, Destrade, Duffy, et al., 2017b; Galbraith et al., 1993; Gallant, 1992; Geddes et al., 2003; Smith et al., 1999; Y. P. Zhang et al., 2014). Estimating the strain and stress distributions in damaged nerve fibres and bundles is a key issue both for clinical care and medical devices (Hemphill et al., 2015; Ma et al., 2016; Shi & Pryor, 2002; Wright & Ramesh, 2012).

Here, two cases of interest provide insights into the electrophysiological impairments of axonal injury due to sudden trauma-induced loading conditions. The boundary conditions in this study replicate the experiments conducted on nerve bundles and axons under both pressure (Hosmane et al., 2011) and elongation (A.C. Bain, Raghupathi, & Meaney, 2001; Geddes et al., 2003; Smith et al., 1999). Additionally, the use of a 3D geometry highlights the difference in voltage and strain distributions in unmyelinated and myelinated fibres in bundles of different size.

Differences in signal transmission arise in the bundle for each fibre, depending on the fibre type. In the bundle, Fibre #3 is activated by imposing a voltage Gaussian distribution on the fibre, while the other fibres are activated based on the voltage gradient from the active fibre and total strains (elastic, equivalent thermal and plastic strains) read at the nerve membrane.

The inclusion of plasticity shows the impact of permanent deformation on signal propagation after a mechanical load is applied. Permanent deformations occur if the strains at the nerve membrane are higher than 21 % according to Bain and Meaney (Allison C Bain & Meaney, 2000), so the reversal voltage potentials change permanently accordingly to the strain intensity. In the cases considered here, the signal read at the nerve membrane on Fibre #3 varies between ranges of voltage value lower than the action potential described by Hodgkin and Huxley (Hodgkin & Huxley, 1952), or the membrane potential is about the baseline value. This means that the distribution of voltage in the bundle is changed, and the other fibres are not activated, because the voltage read at their nerve membranes is a subthreshold stimulation, so that an action potential cannot be elicited, see Figure 5.

During compression, the neural activity is changed according to the elastic strain at the nerve membrane, where we find plastic strains of 0.5 % in unmyelinated fibres, but no plastic strains in myelinated fibres for the range of pressure levels applied. Although we chose high pressure values comparable to those found in TBI (Hosmane et al., 2011), the applied strains do not compromise the functionality of the membrane and its ability to generate signals, even when assuming changes in ionic conductance (Boucher et al., 2012). Additionally, the resulting total strains at the membrane are lower than 3.5 % and 1.5 % in unmyelinated and myelinated bundles. Thus, during compression, the fibre is far from mechanical failure thanks to the small strains generated at the membrane. Fibres and bundles appear to be stronger in compression than in elongation. A uniform compression of the bundle induces only a 1% elongation of the bundle, leading to small changes in neural activity and lower values of plastic strains ($PEs$) are found at the membrane.

During elongation, results show that the neural activity is more easily affected by deformations in small bundles than in larger bundles, where at 14% of elongation a plateau indicates a new osmotic gradient across the nerve membrane, see Figure 5. As in experiments, the larger the fibre, the higher the voltage read at the membrane (Durand, 2000; Galbraith et al., 1993; Smith et al., 1999), and hence the higher the deformations, see Figure 7, whose plastic component is shown in Figure 6. We also find that the myelin layer constrains the mechanical deformation of the nerve membrane at the Ranvier nodes, generating a different distribution of plastic strain around the fibre. This important property of myelin preserves the functionality of the membrane by distributing the applied uniaxial strain within the bundle. It is revealed thanks to 3D Finite Element modelling.

Small myelinated bundles tolerate lower plastic strain than unmyelinated bundles, as seen in previous studies (Hemphill et al., 2015; Jafari et al., 1998; Reeves et al., 2012). However, smaller plastic strains are found in bigger myelinated bundles, where plastic damage occurs locally around the fibre rather than along the fibre length, see Figure 5. This could be thought of as a way to preserve good communication between neurons cells under stretch. The nervous cell reading the signals carried by a damaged myelin fibre might not consider it as a valuable source of information because of its reduced magnitude. For higher deformations, the action potential is not elicited because of the very low ionic gradient across the nerve membrane (Cinelli, Destrade, Duffy, & McHugh, 2017a), altering the signal propagation from cell-to-cell, and in turn, the communication between cells.

The assumption of instantaneous loading is a first step towards the electro-mechanical analysis of changes associated with TBI. Rate-dependent loadings could be included in future works. Injury pathologies in nerve fibres are also initiated and influenced by strain and strain rate, which have a significant impact on the time of neural death and pathomorphology, respectively (Bar-kochba et al., 2016). Experimental studies on human axons show morphological changes of axons at different stages of dynamic stretch injury (Smith et al., 1999). Axons can tolerate stretching up to twice their original length under slow loading rates (within the range of minutes (Tang-schomer et al., 2017)), with elastic recovery of the initial pre-stretched geometry (Bar-kochba et al., 2016; Tang-schomer et al., 2017). However, dynamic loading conditions with a short pulse duration (lower than $50 s^{-1}$ (Tang-schomer et al., 2017)), initiate undulating distortions along their entire length (Bar-kochba et al., 2016; Smith et al., 1999; Tang-schomer et al., 2017), and recovery of the pre-stretched geometry was found to be non-uniform. Axonal regions can manifest both an elastic recovery and a delayed elastic response, i.e. a gradual recovery, along the same fibre length (Smith et al., 1999). For example, mechanical failure of squid giant axons was found at $25 - 30\%$ stretch at a strain rate of $10 s^{-1}$ (Galbraith et al., 1993), while human axons, with a diameter of about $0.5 - 1$ $\mu m$, tolerate dynamic stretch injury at strains up to 65% according to Smith et al.(Smith et al., 1999).

Finally, as highlighted earlier, our model assumed an idealized geometry of a nerve bundle. Further works must tackle the effect of a realistic geometry of nervous cells, by considering fibre alignments in different directions and multiple fibre activation.

IV.  Conclusion

We propose a fully coupled electro-mechanical framework for modelling the biophysical phenomena accompanying neural activity, such as electrostriction and piezoelectricity, by relying on the electro-thermal analogy. This framework is a new approach in neural engineering, embracing the main findings of experimental observations.  The model, built on previously published work (Cinelli, Destrade, Duffy, et al., 2017a, 2017b, 2017c) incorporates the effect of plasticity to generate a fully coupled 3-dimensional simulation of ion channel leaking for nerve fibres under pressure and displacement loads. To recapitulate, this model shows that:

- Time-shifted, signal magnitude and nerve membrane potential baseline values are found to be dependent on the total strain, voltage and size of the fibre;
- Lower strain and lower electrophysiological changes are found in myelinated fibres than in unmyelinated fibres;
- The myelin layer redistributes the generated plastic strain within the bundle;
- Fibres and bundles are stronger under compression than elongation;
- During elongation, mechanical failure occurs at lower elongation in BBMY, than in BBUN, SBMY and SBUN;
- Larger bundles deform more than small bundles;
- Larger bundles fail because of elastic strain, not plastic strain;
- Trauma affects small bundles more than larger bundles;
- Plastic strains are not influenced by the trauma level at the nerve membrane (as measured by $AC$);
- Trauma ($AC$) does not influence the membrane baseline voltage during compression or elongation;
- Trauma does not influence voltage and plastic strain in larger bundles.

This model can contribute to the understanding of the causes and consequences of traumatic brain injury and diffuse axonal injury to improve diagnosis, clinical treatments and prognosis by simulating the mechanical changes accompanying the changes in signal transmission in TAI-induced loading conditions.

Acknowledgements

We gratefully acknowledge funding from the Galway University Foundation, the Biomechanics Research Centre and the Power Electronics Research Centre, College of Engineering and Informatics at NUI Galway.